\documentclass[prb,aps,twocolumn,amsmath,amssymb,floatfix,
superscriptaddress]{revtex4}

\usepackage[dvips]{graphics}
\usepackage{color}
\definecolor{dred}{rgb}{0.75,0,0}

\begin{document}

\title{Phase controlled metal-insulator transition in multi-leg 
quasiperiodic optical lattices}

\author{Santanu K. Maiti$^{*,}$}

\affiliation{Physics and Applied Mathematics Unit, Indian Statistical
Institute, 203 Barrackpore Trunk Road, Kolkata-700 108, India}

\author{Shreekantha Sil}

\affiliation{Department of Physics, Visva-Bharati, Santiniketan, West
Bengal-731 235, India}

\author{Arunava Chakrabarti}

\affiliation{Department of Physics, University of Kalyani, Kalyani,
West Bengal-741 235, India}

\begin{abstract}

A tight-binding model of a multi-leg ladder network with a continuous 
quasiperiodic modulation in both the site potential and the inter-arm 
hopping integral is considered. The model mimics optical lattices where 
ultra-cold fermionic or bosonic atoms are trapped in double well potentials.  
It is observed that, the relative phase difference between the on-site 
potential and the inter-arm hopping integral, which can be controlled by 
the tuning of the interfering laser beams trapping the cold atoms, can 
result in a mixed spectrum of one or more absolutely continuous subband(s) 
and point like spectral measures. This opens up the possibility of a 
re-entrant metal-insulator transition. The subtle role played by the relative 
phase difference mentioned above is revealed, and we corroborate it 
numerically by working out the multi-channel electronic transmission for 
finite two-, and three-leg ladder networks. The extension of the calculation 
beyond the two-leg case is trivial, and is discussed in the work.

\end{abstract}

\maketitle

\section{Introduction}

Optical lattices loaded with ultra-cold bosonic or fermionic quantum gases 
have dominated the research in low-dimensional systems in the past 
decade~\cite{jak,bloch,lewen,dutta1,dutta2}. Novel, synthetic lattices have  
been designed with quantum degenerate gases taking advantage of an 
unprecedented control over the interfering laser beams which trap such 
atoms in tailor made geometries. The achievement of a fine tuning of the 
lattice depth, geometry or filling factor, has paved the way for direct 
experimental observation such as the exotic superfluid to Mott insulator 
transition~\cite{greiner}, or, the  generation of synthetic magnetic 
fields to realize the Hofstadter butterfly in ultra-cold gases loaded in 
a 2D optical lattice~\cite{atala,miyake,gsun}.

Quasiperiodic potentials typically modeled by the well known 
Aubry-Andr\'{e}-Harper (AAH) profile~\cite{aubry} have been famous for 
theoretically 
predicting (in parameter space) a phase transition in the character of 
single particle states. Recent success in engineering such potentials in 
photonic crystals~\cite{negro, lahini, kraus}, and in optical lattices of 
ultra-cold atoms~\cite{roati,modugno} has brought this model alive with 
deeper prospects of studying quasiperiodic systems from an experimental 
perspective, as well as for exploring emerging topological phases of matter. 
In addition, Anderson localization~\cite{anderson} of matter waves has 
been experimentally observed for the first time in such quasiperiodic 
optical lattices~\cite{roati}.

In this communication we propose a multi-leg, quasiperiodic optical lattice, 
with AAH modulation given both in the distribution of the on-site potentials 
and the inter-arm hopping integrals. This is shown to serve 
as a model system where one can engineer practically a continuous 
change in the transport properties of the network, going from  a low (almost 
zero) value to a high transport (metallic) regime as a function of the Fermi 
energy, being lowered again beyond certain range in the energy eigenvalues. 
This points out to the possibility of observing a re-entrant transition. 
The engineering of the energy bands is accomplished by a decoupling of the 
Schr\"{o}dinger equation, cast in the form of difference equations, which 
initially involve a coupling between the legs through a tunnel hopping 
integral, by exercising a uniform change of basis. We provide an analytical 
\begin{figure}[ht]
{\centering \resizebox*{7.5cm}{4cm}{\includegraphics{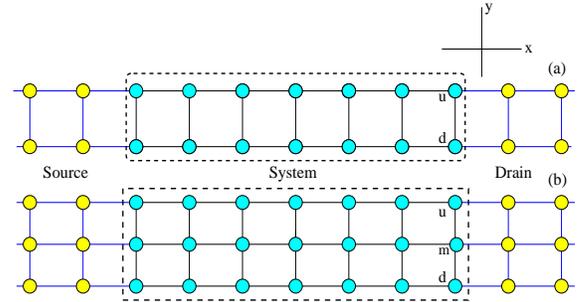}}\par}
\caption{(Color online). Schematic view of (a) two-leg and (b) 
three-leg ladder networks (cyan colored sites) connected to semi-infinite 
multi-channel source and drain (yellow colored sites). The upper and the 
lower legs are marked `u' and `d' respectively for the two-leg ladder 
and `u', `m' and `d' for the three-leg one.}
\label{ladder}
\end{figure}
method for this, and use a two-leg ladder as a prototype case for 
illustration. We prove that, this effect can be observed by tuning the 
{\it relative phase difference} of the lattice potential and the 
inter-arm hopping integrals. This is interesting as 
the `phases' can now be controlled experimentally. The individual values 
of such phases do not not affect the bulk properties of a one-dimensional 
AAH model, preserving its self-duality, and yet, they are shown to exhibit 
nontrivial topological properties that are attributed to a higher dimensional 
system~\cite{kraus}. In addition, the quasiperiodic modulation in the hopping 
integrals in the commensurate limit turns out to be topologically nontrivial, 
supporting zero energy edge modes~\cite{sankar1}. 

It is pertinent to mention that, multi-leg ladders have become relevant 
recently after the experimental realization of double well optical lattices 
where bosonic atoms ($^{87}$Rb) are trapped~\cite{sebby}. A one-dimensional 
double well optical lattice is equivalent to a two legged ladder 
network~\cite{danshita}. Construction of such systems exploits the flexibility 
and control over the interfering laser beams, and trapping of bosonic or 
fermionic ultra-cold atoms leading to the realization of two and three legged 
ladders~\cite{piraud,greschner,hugel,kolley}. 

The result obtained here is in sharp contrast to a purely one-dimensional 
AAH model where the state transition occurs depending on the strength of 
the potential only, that is, in a parameter space rather than as a function 
of the Fermi energy, which is the case here. We present analytically exact 
results in details for a two-leg optical ladder lattice, and discuss 
the pathway to generalize this to multiple legs, leading finally to a 
two-dimensional structure. 

The rest of the paper is arranged as follows. In the first part of Sec. II 
we present analytically exact results and in the second part of this
section we give the theoretical prescription to simulate a real experimental
situation by coupling the ladder to ideal semi-infinite leads. The numerical
results are placed in Sec. III and finally we conclude in Sec. IV.

\section{Analytical argument and theoretical framework}

\noindent
{\bf Preservation of duality and spectral character in one-dimension}
\vskip 0.5cm
\noindent
Before going into the actual problem addressed here, it is important to 
examine if the presence of a phase factor does, in any way, influence 
the spectral character of a one-dimensional AAH model. In its simplest 
form the on-site potential of an AAH model in a one-dimensional lattice 
of lattice constant $a$ is written as, $\epsilon_n = 
\lambda \cos(2\pi Q n a +\varphi)$, where $Q$ is an irrational number, 
typically chosen as the golden mean $(\sqrt{5}+1)/2$ in what is 
called the {\it incommensurate limit}. $\lambda$ is the strength 
of the on-site potential and $\varphi$ is an arbitrary phase factor that 
shifts the origin of the potential. The Schr\"{o}dinger equation, written 
in the form of a difference equation, relating the amplitudes $f_n$ on 
the neighboring sites on the lattice is given by, 
\begin{equation}
[E - \lambda \cos(2\pi Q n a + \varphi)] f_n = t (f_{n-1} + f_{n+1})
\label{diff1d}
\end{equation}
where, $t$ is the nearest-neighbor hopping integral. A discrete Fourier 
transform, $f_n = \exp(i\varphi n a) \sum_{m} 
g_m \exp[i m (2 \pi Q n a + \varphi)]$ maps Eq.~\ref{diff1d} into
\begin{equation}
[E - t \cos(2\pi Q m a + \varphi)] g_m = (\lambda/2) (g_{m-1} + g_{m+1})
\label{diff1d2}
\end{equation}
The AAH model leads to the conclusion that, if $g_m$ is localized, that is, 
if $\sum_{m=-\infty}^{\infty} |g_m|^2$ is finite, then 
$f_n = \exp(i \varphi n a) g_m \exp[i m (2\pi Q n a + \varphi)]$ represents 
an extended wave-function, i.e., $\sum_{n=-\infty}^{\infty} |f_n|^2$ diverges, 
and that, if $f_n$ is localized, then $g_m = \exp(i\varphi m a) \sum_{n}
f_n \exp[i n (2 \pi Q m a + \varphi)]$ will be extended in character. 
This is the essence of duality~\cite{sokoloff}. Eq.~\eqref{diff1d2} thus 
tells us that the `duality' is preserved even if we add an arbitrary phase 
to the potential.

With the assumed form of the on-site potential, viz, 
$\epsilon_n = \lambda \cos(2\pi Q n a + \varphi)$, it now becomes a simple 
task to show that in the Fourier space the phase can be absorbed in the 
amplitude of the nearest neighbor hopping integral, making the effective 
on-site potential free of phase. The difference equation reads (in Fourier 
space), 
\begin{equation}
\left[E - t \cos(2 \pi Q m a)\right] g_m = (\lambda/2) e^{-i\varphi} g_{m+1} 
+ (\lambda/2) e^{i\varphi} g_{m-1}
\label{diff1d3}
\end{equation}
It is obvious that, the effective hopping integrals in the Fourier space 
are $t^{eff}_{m, m \mp 1} \equiv (\lambda/2) \exp(\pm i\varphi)$. Since 
both the effective potential $\epsilon^{eff}_{m} \equiv t \cos(2 \pi Q m a)$ 
and the magnitude of the nearest-neighbor hopping integral 
$|t^{eff}_{m, m \pm 1}|$ do not depend on the phase $\varphi$, the density 
of states and the localization properties are in no way, affected by any 
variation of this phase. That is, we still retain the well known AAH 
conclusion that the single particle states will be extended, localized or 
critical if $\lambda$ is less than, greater than, or equal to $2t$ 
respectively.

In what follows we show that the situation is not so trivial in quasi 
one-dimensional cases, and with phases introduced in both the potential 
term and the hopping integrals along the rungs.

\vskip 0.5cm
\noindent
{\bf The case of a multi-leg ladder network: The metal-insulator 
transition}
\vskip 0.5cm
\noindent
The model is schematically shown in Fig.~\ref{ladder}, which mimics a 
two-leg optical lattice (a), and a three-leg one (b). The cyan colored 
sites represent the system of atoms loaded in the up (u) and down (d) 
legs. The rungs are along the $y$-direction. The leads needed for the 
transport calculation are shown by yellow dots. For a ladder extending to 
infinity along the $x$-direction, the Hamiltonian for the system (S), 
written in a tight-binding framework reads,  
\begin{eqnarray}
\mbox{\boldmath $H_S$} &=& \sum_n \mbox{\boldmath $c_n^{\dagger} 
\epsilon_n c_n$} + \sum_n \left(\mbox{\boldmath $c_{n+1}^{\dagger} t$} 
\mbox {\boldmath $c_n$} + h.c. \right) 
\label{ham}
\end{eqnarray}
where, $n=-\infty ...-1$, $0$, $1$, $2$, $\dots$, $\infty$ is the site 
index along the $x$-direction. The operators in Eq.~\ref{ham} are as 
follows.\\
\mbox{\boldmath $c_n^{\dagger}$}=$\left(\begin{array}{cc}
c_{n,u}^{\dagger} & c_{n,d}^{\dagger} 
\end{array}\right);$
\mbox{\boldmath $c_n$}=$\left(\begin{array}{c}
c_{n,u} \\
c_{n,d}\end{array}\right);$
\mbox{\boldmath $\epsilon_n$}=$\left(\begin{array}{cc}
\epsilon_{n,u} & t_{n,y} \\
t_{n,y} & \epsilon_{n,d} \end{array}\right);$ 
\mbox{\boldmath $t$}=$\left(\begin{array}{cc}
t_x & 0 \\
0 & t_x \end{array}\right).$ \\
\vskip 0.15cm
\noindent
Here $\epsilon_{n,u(d)}$ is the on-site potential at the $n$-th site of the 
`up' (u) or `down' (d) leg. We choose 
$\epsilon_{n,u} = \epsilon_{n,d} = \lambda \cos(2\pi Q n + \varphi_\nu) 
= \epsilon_n$ (say). The lattice constant $a$ is chosen as unity.
$t_{n,y}$ is quasiperiodically modulated hopping integral along the $n$-th 
rung and is chosen as, $t_{n,y}= t_y \cos(2\pi Q n + \varphi_\beta)$.
$Q$ will be chosen as the golden ratio $(\sqrt{5}+1)/2$, whose rational 
approximants have been exploited to explore the topological states in the 
1D version of the lattice~\cite{sankar1}. 
$c_{n,u(d)}^{\dagger}$ ($c_{n,u(d)}$) is the creation (annihilation) operator 
of a particle at the $n$-th site in the up or down arm of the ladder. We have 
kept the hopping integral along the $x$-direction constant, and equal to 
$t_x$. A variation in $t_x$ can easily be dealt with in the present formalism.

Let us explain the scheme by analyzing the spectrum of a two legged ladder 
of infinite extension along the $x$-direction. The Schr\"{o}dinger equation 
for such a ladder network can equivalently be written down in the form of 
a couple of difference equations, viz.,
\begin{eqnarray}
(E - \epsilon_{n,u}) \psi_{n,u} & = & t_x (\psi_{n+1,u} + \psi_{n-1,u}) + 
t_{n,y} \psi_{n,d}  \nonumber \\
(E - \epsilon_{n,d}) \psi_{n,d} & = & t_x (\psi_{n+1,d} + \psi_{n-1,d})+ 
t_{n,y} \psi_{n,u}  
\label{diff}
\end{eqnarray}
We have chosen to set, for our purpose, $\epsilon_{n,u} = \epsilon_{n,d} 
= \epsilon_n$.
In that case, it is easily seen that, the potential and the hopping matrices 
are nothing but, $\mbox{\boldmath $\epsilon_n$}=\epsilon_n \mbox{\boldmath $I$} 
+ t_{n,y} \mbox{\boldmath $\sigma_x$}$, and $\mbox{\boldmath $t$} = 
t_x \mbox{\boldmath $I$}$, where, \mbox{\boldmath $\sigma_x$} and 
\mbox{\boldmath $I$} are the Pauli matrix and the $2 \times 2$ identity 
matrix respectively. It now becomes obvious that, the commutator 
$[\mbox{\boldmath $\epsilon_n$}, \mbox{\boldmath $t$}] = 0$ independent of the 
rung index `$n$'. This allows us to diagonalize the potential and the hopping 
matrices simultaneously and make a {\it uniform change of basis}, going from 
the Wannier orbitals $ \mbox{\boldmath $\Psi_n$} 
\equiv (\psi_{n,u}, \psi_{n,d}) $ to $ \mbox{\boldmath $\Phi_n$} \equiv 
(\phi_{n,u}, \phi_{n,d}) = \mbox{\boldmath $M^{-1}$} 
\mbox{\boldmath $\Psi_n$}$. 

The change of basis decouples the set of Eq.~\eqref{diff} as, 
\begin{eqnarray}
\left [ E - \lambda_1 \cos (2 \pi Q n + \xi_1) \right ] \phi_{n,u} & = & 
t_x (\phi_{n+1,u} + \phi_{n-1,u}) \nonumber \\
\left [ E - \lambda_2 \cos (2 \pi Q n + \xi_2) \right ] \phi_{n,d} & = & 
t_x (\phi_{n+1,d} + \phi_{n-1,d}) \nonumber \\
\label{decouple}
\end{eqnarray}
where, \\
$\lambda_1 = \sqrt{\lambda^2 + t_y^2 + 2 \lambda t_y 
\cos(\phi_\nu - \phi_\beta)}, \\
\lambda_2 = \sqrt{\lambda^2 + t_y^2 - 2 \lambda t_y 
\cos(\phi_\nu - \phi_\beta)}, \\
\xi_1 = \tan^{-1}[(\lambda \sin \phi_\nu + t_y \sin \phi_\beta)/
(\lambda \cos \phi_\nu + t_y \cos \phi_\beta)], \\
\xi_2 = \tan^{-1}[(\lambda \sin \phi_\nu - t_y \sin \phi_\beta)/
(\lambda \cos \phi_\nu - t_y \cos \phi_\beta)]$.\\
\vskip 0.01cm
\noindent
The new states are,
\begin{eqnarray}
\phi_{n,u} &=& (-1/2) \psi_{n,u} + (1/2) \psi_{n,d} \nonumber \\
\phi_{n,d} &=&  (1/2) \psi_{n,u} + (1/2) \psi_{n,d}
\label{pseudo}
\end{eqnarray}
and it should be appreciated that $\phi_{n,u(d)}$ can be localized 
only when both $\psi_{n,u(d)}$ are localized. If any of the original 
Wannier orbital $\psi_{n,u(d)}$ corresponds to an extended eigenstate, 
then the {\it new} state $\phi_{n,u}$ will retain its extended character.

It is apparent that, the first (or the second) of the set of 
Eq.~\eqref{decouple} will give extended, localized or critical eigenstates 
for $\lambda_1$ (or $\lambda_2$) $<$, $>$ or $= 2t_x$ ~\cite{aubry}. This 
implies that, if we set for example, $\lambda_1 < 2t_x$, the first of the 
Eq.~\eqref{decouple} will yield an absolutely continuous spectrum, while 
the second one can give rise to a pure point, critical or even absolutely 
continuous energy spectrum depending on the magnitude of $\lambda_2$ with 
the choice of parameters prefixed for $\lambda_1$. Thus we have an option of 
engineering the spectrum. As the density of states of the full quasi-one 
dimensional ladder will come from a convolution of the two densities of 
states obtained from the two decoupled equations, we have the provision of 
engineering the full density of states, creating even a mixed one with 
localized and extended eigenfunctions. The possibility of a re-entrant 
metal to insulator crossover is thus on the cards. 

\begin{figure}[ht]
{\centering \resizebox*{5.25cm}{4cm}{\includegraphics{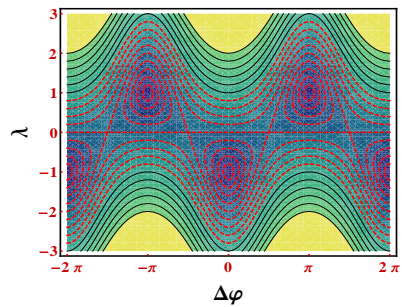}}\par}
\caption{(Color online). Variation of the strength of the effective potential 
$\lambda_1$ as a function of the original strength of potential $\lambda$ and 
the relative phase difference $\Delta\varphi = \phi_\nu - \phi_\beta$. We 
have set $t_x=1$ and $t_y=1$. The red dotted contours provide combinations 
of $(\Delta\varphi, \lambda)$ for which the values of $\lambda_1$ is less 
than $2$.} 
\label{contour}
\end{figure}

In Fig.~\ref{contour} we have shown the contour plot of the strength of the 
on-site potential $\lambda_1$ as the relative phase difference 
$\Delta\varphi = \varphi_\nu - \varphi_\beta$ is varied against a wide 
choice of the basic potential strength $\lambda$. We have set the amplitude 
of the transverse hopping $t_y=1$. The contours correspond to various values 
of $\lambda_1$ less than, or greater than $2t_x=2$ (with $t_x$ being set 
equal to unity) depending on the combination $(\lambda$, $\Delta\varphi)$. 
The observation is similar for $\lambda_2$ as well. If one dwells in that 
regime of the parameter subspace  $(\lambda_{1(2)}$, $\Delta\varphi)$, 
where $\lambda_1$ is less than $2t_x$, (red dotted contours) while 
$\lambda_2$ can be tuned (by $\Delta\varphi$) to be  greater than $2t_x$, 
then the first of the decoupled equations Eq.~\eqref{decouple} yields an 
absolutely continuous spectrum, while the second one gives rise to a pure 
point spectrum. The energy spectrum of the two arm ladder network is a 
mixed one then, as discussed before, comprising an absolutely continuous part 
populated by Bloch like extended states and exponentially localized states 
arising out of the second of the Eq.~\eqref{decouple}. The proximity of the 
edges of the {\it extended} band and the subclusters formed by the 
{\it localized} eigenfunctions will determine the possibility of a 
metal-insulator transition. This is sensitive to the appropriate choice of 
the phase difference $\Delta\varphi$ and $\lambda$. 

To check the truthfulness of the above analysis, we consider 
non-interacting, spinless electrons on a two legged ladder network.
For this system, we evaluate the density of states and multi-channel 
transport with an AAH variation in both the potentials $\epsilon_n$ and 
the inter-arm hopping integral $t_y$, as given in the 
previous discussion and including the phases. To simulate a real 
experimental situation we clamp an $N$-rung two-legged ladder at its 
extremities to ideal semi-infinite `electrodes' (leads) of finite width, 
and evaluate the average density of states of a finite sized system with 
the leads attached. The Hamiltonian of the electrodes is taken to be 
\mbox{\boldmath $H_0$} $+$ \mbox{\boldmath $H_{LS}$} $+$ 
\mbox{\boldmath $H_{RS}$} where, 
\begin{eqnarray}
\mbox{\boldmath $H_0$} &=& \sum_{n=-\infty}^{0} \mbox{\boldmath $c_n^{\dagger}
\epsilon_0 c_n$} + \sum_{n=N+1}^{\infty} \mbox{\boldmath $c_n^{\dagger}
\epsilon_0 c_n$} \nonumber \\
 & & +  \sum_n \left(\mbox{\boldmath $c_{n+1}^{\dagger} t_0$}
\mbox {\boldmath $c_n$} + h.c. \right)
\label{hamlead}
\end{eqnarray}
with, 
\mbox{\boldmath $\epsilon_0$}=$\left(\begin{array}{cc}
\epsilon_{0} & \tau_{y} \\
\tau_{y} & \epsilon_{0} \end{array}\right)$ and
\mbox{\boldmath $t_0$}=$\left(\begin{array}{cc}
\tau_x & 0 \\
0 & \tau_x \end{array}\right).$ \\
$\tau_x$ and $\tau_y$ are the hopping integrals in the leads, 
along the $x$ and the $y$-directions respectively.
The couplings between the left (L) and the right (R) 
electrodes, and the system (S) are given by, 
$\mbox{\boldmath $H_{LS}$} = \mbox{\boldmath $c_0^{\dagger}$} 
\mbox{\boldmath $t_L$}\mbox{\boldmath $c_1^{\dagger}$}$, and 
$\mbox{\boldmath $H_{RS}$} = \mbox{\boldmath $c_{N}^{\dagger}$}                       
\mbox{\boldmath $t_R$}\mbox{\boldmath $c_{N+1}^{\dagger}$}$ respectively, 
with, 
\mbox{\boldmath $t_L(R)$}=$\left(\begin{array}{cc}
\tau_{L(R)S} & 0 \\
0 & \tau_{L(R)S} \end{array}\right) $.
The average density of states of the electrode-system-electrode network is 
given, in Green's function formalism as, 
\begin{equation}
\rho_{av} = -\frac{1}{N \pi} \mbox{Im} [\mbox{Tr}\,\mbox{\boldmath $G$}]
\end{equation}
where, $\mbox{\boldmath $G$} = [(E + i \eta) \mbox{\boldmath $I$} - 
\mbox{\boldmath $H$} ]^{-1}$, 
The full Hamiltonian $\mbox{\boldmath $H$} = \mbox{\boldmath $H_{LS}$} 
+ \mbox{\boldmath $H_S$} + \mbox{\boldmath $H_{RS}$}$.
In terms of the Green's function of the system and the electrode-system 
couplings, the transmission coefficient across the system is written 
as~\cite{datta}
\begin{equation}
T = \mbox{Tr}[\mbox{\boldmath $\Gamma_L$} \mbox{\boldmath $G_S^{r}$} 
\mbox{\boldmath $\Gamma_R$} \mbox{\boldmath $G_S^{a}$}]
\label{trans}
\end{equation}
Here, \mbox{\boldmath $G_S^{r}$} and \mbox{\boldmath $G_S^{a}$} are the 
`retarded' and the `advanced' Green's functions respectively, of the 
two leg ladder network, including the effects of the leads. The Green's 
function for the system concerned is expressed as, 
\begin{equation}
\mbox{\boldmath $G_S$} = (E \mbox{\boldmath $I$} - \mbox{\boldmath $H_S$} 
- \mbox{\boldmath $\Sigma_L$} - \mbox{\boldmath $\Sigma_R$} )^{-1}
\label{selfenergy}
\end{equation}
where, \mbox{\boldmath $\Sigma_{L(R)}$} are the self energies and take 
care of the system-electrode coupling~\cite{datta}.

\section{Numerical results}

\begin{figure}[ht]
{\centering \resizebox*{8cm}{7cm}{\includegraphics{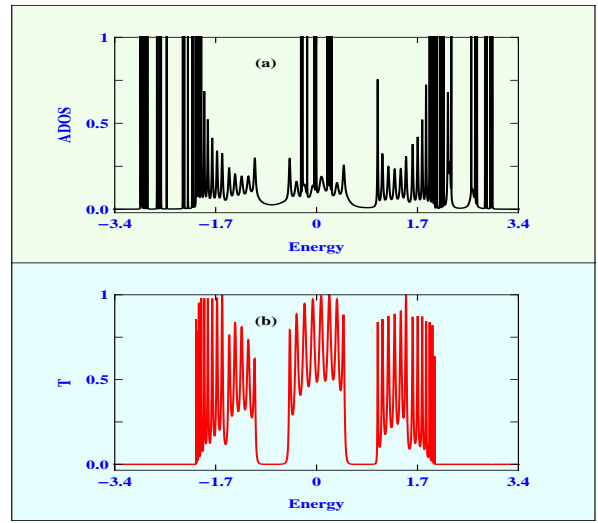}}\par}
\caption{(Color online). (a) Density of states and (b) the transmission 
characteristics of the two arm aperiodic ladder network with $40$ rungs. 
Here, $\varphi_\nu = \varphi_\beta = 0$, $\lambda=1.5$, $t_x=t_y=1$, 
$\tau_x=\tau_y=2$, $t_{L(R)}=1$ and $Q=(\sqrt{5}+1)/2$. 
The chemical potential is set at zero.}
\label{phasezero}
\end{figure}

\begin{figure}[ht]
{\centering \resizebox*{8cm}{7cm}{\includegraphics{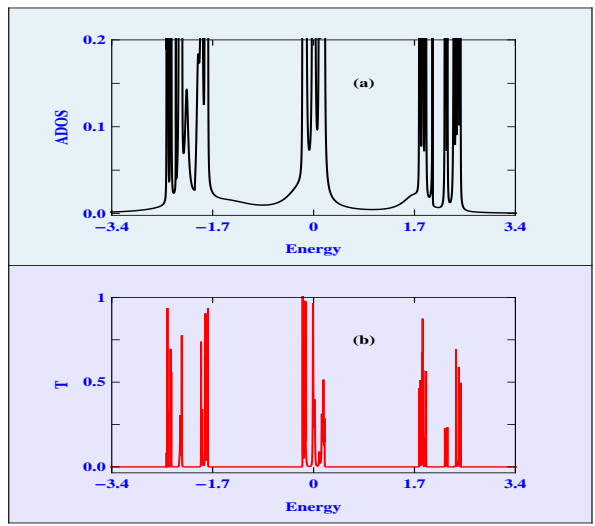}}\par}
\caption{(Color online). (a) Density of states and (b) the transmission 
characteristics of the two arm aperiodic ladder network with $40$ rungs 
when $\phi_\nu = 0$ and $\phi_\beta = \pi/2$. We take 
$\lambda_1=\lambda_2=\sqrt{3.25}$, and the values of other parameters
are same as taken in Fig.~\ref{phasezero}.}
\label{phasepiby2}
\end{figure}

In Fig.~\ref{phasezero} and Fig.~\ref{phasepiby2} we show the density of 
states (panel (a)) and the transmission coefficient (panel (b)) for a 
$40$-rung ladder network with $\varphi_\nu = \varphi_\beta =0$, and 
$\varphi_\nu = 0$, $\varphi_\beta = \pi/2$ respectively. We have set 
$\lambda=1.5$, $t_y=1$, $Q=(\sqrt{5}+1)/2$ and $t_x=1$. The imaginary part 
added to the energy, viz. $\eta$ has been taken to be $2 \times 10^{-9}$, 
and the chemical potential has been set equal to zero.
With the above choice of parameters, and when $\phi_\nu = \phi_\beta =0$, 
we find $\lambda_1=2.5$ which is greater than $2t_x=2$, and hence the first 
set of equations Eq.~\eqref{decouple} yields a localized spectrum. However, 
$\lambda_2=0.5$ in this case, and is less than $2t_x=2$, and the second of 
the set of Eq.~\eqref{decouple} yield an absolutely continuous spectrum 
populated by extended states only. In Fig.~\ref{phasezero} we thus observe a 
mixed spectrum, the extended zone $-2<E<2$ destroying the localized character 
of the eigenstates arising out of the second equation. The localized states 
which lie outside this `extended state zone' appear as spikes, densely 
packed towards the flanks. The transmission coefficient corresponding to 
the extended eigenstates residing in the absolutely continuous parts of the 
spectrum turns out to be high and corroborates the conclusion about the 
character of the eigenfunctions.

In contrary to this, in Fig.~\ref{phasepiby2}, both $\lambda_1$ and 
$\lambda_2$ turn out to be equal to $\sqrt{3.25}$ which is less that 
$2t_x=2$. Only {\it extended states} are generated by each of the decouple 
equations in Eq.~\eqref{decouple}. Thus the full spectrum 
of the two-leg ladder consists of extended eigenstates only. This is 
precisely corroborated by the transmission characteristics shown in 
Fig.~\ref{phasepiby2}, which show clusters of finite transmission exactly 
at the place where the density of states turns out to be non-zero. The 
system now, is totally metallic. Thus, just by tuning the relative 
phases one can engineer whether to have a metallic system out of the 
ladder network or to create a mixed spectrum for the same where one can 
see a non-metallic (insulating) and metallic (conducting) character 
depending on the position of the Fermi level.

Getting back to the discussion of Fig.~\ref{phasezero},the 
prime interest lies in the regions in between the subclusters of extended 
states around $E = \pm 0.85$ where the density of states in low, but finite 
and remains finite even if we lower the value of the imaginary part $\eta$ 
further. We have scanned this region by decreasing the 
energy interval gradually. The states occupying this zone in the energy 
scale appear as sharp peak (the peak structure is not visible at this 
scale) and form a dense cluster that appears almost as continuous. This 
is concluded from finer and finer scanning of the energy interval around 
$E = \pm 0.85$. The corresponding transmission for all such states is zero. 
This signifies that these are sub-bands of exponentially localized 
eigenstates.

On close scrutiny, done numerically, it also appears that 
such {\em minibands} of localized eigenstates merge with the neighboring 
absolutely continuous bands of extended states in a continuous manner i.e., 
we were unable to detect any clear separation between the end of the 
band of localized states and the beginning of the band of extended 
eigenstates. This happens also, in the same manner, as one crosses the band 
of high transmission and re-enter into the next miniband of localized 
states. This tempts us to conclude that a re-entrant metal-insulator 
transition is definitely a 
possibility here, as anticipated from the exact analytical treatment 
presented before. The same kind of behavior is observed when we have 
chosen $\Delta\varphi=\pi/2$, as shown in Fig.~\ref{phasepiby2}. 
Here, $\lambda_1 = \lambda_2 = \sqrt{3.25}$ and less than 
$2t_x$. The non-zero DOS regimes and their corresponding zero transmission 
values strengthen our surmise that, a re-entrant metal-insulator transition 
can indeed be triggered by a variation of the phase difference 
$\Delta\varphi = \varphi_\nu - \varphi_\beta$. 

It would be interesting to link such observations to experiments with 
Bose-Einstein condensates released into one-dimensional waveguides in 
the presence of a controlled disorder~\cite{roati, billy}. 
A typical method to extract information about localization has been to 
photograph the atomic density profiles~\cite{roati,billy} as a function 
of time. Disorder is found to stop the expansion of the wave packet 
and lead to the formation of exponentially localized wave 
function~\cite{roati,billy}. In a very recent experiment Schreiber 
{\em et al.}~\cite{schreiber} made a 
direct observation of quantum many body localization in interacting 
fermionic systems using $^{40}K$ atoms in a trap of an AAH potential 
profile with a phase offset. They have measured the `imbalance' 
$I = (N_e - N_o)/(N_e + N_o)$ between the occupation of the `odd' (o) and 
`even' (e) sites in the optical lattice. $I$ essentially measures the 
charge density wave (CDW) order which basically acts as an {\it order 
parameter} to characterize localization effects, and incidentally any 
possible localization-de localization transition. For strong localization, 
the particles remain localized to the initial locations they were `released' 
into. The CDW is smeared a little bit. For extended eigenstates, the 
localization length increases and the steady state value of the CDW is 
lowered~\cite{schreiber}. A weak atomic density can result in a negligible 
interaction among the particles and the non-interacting limit can be 
achieved. The theory presented in this communication can thus be taken as 
a proposal for verifying the existence of a metal-insulator like behavior 
\begin{figure}[ht]
{\centering \resizebox*{8cm}{7cm}{\includegraphics{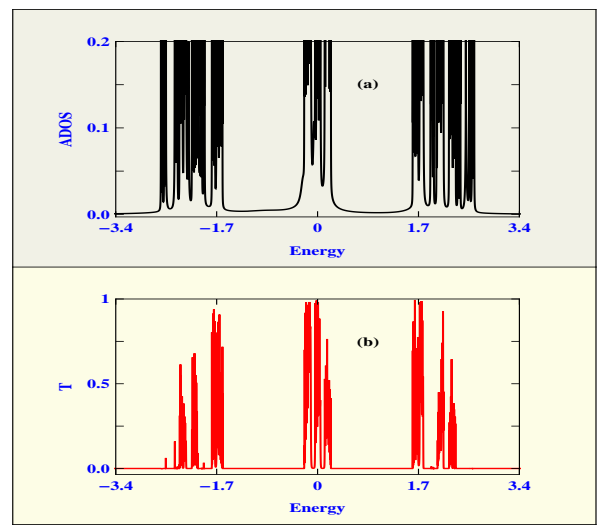}}\par}
\caption{(Color online). (a) Density of states and (b) the transmission 
coefficient of a three arm ladder network with leads of the same 
width attached to it. There are $100$ rungs, and we have chosen 
$\phi_\nu =0$, $\phi_\beta=\pi/2$, $\lambda=1.5$, $t_x = t_y = 1$, 
$\tau_x=\tau_y = 2$ and $t_{L(R)}=1$.} 
\label{threearm}
\end{figure}
in quasi one-dimension and in the non-interacting limit, in the spirit of 
the above experiments. The likely changes in the relaxation dynamics of 
the CDW as mentioned above can bring out the role played by the phase 
difference in regard of the metal-insulator transition.

Extending the above idea to multi-leg ladder networks is trivial. For 
an $N$-leg ladder, the difference equations can be split into a set of 
$N$ decoupled equations. Depending on the relative phase difference, any 
one or more of them can render absolutely continuous spectrum and/or a 
mixing of the same with pure point spectrum, leading to the possibility 
of a metal-insulator transition in quasi one- or even two-dimensional 
quasiperiodic optical lattice models. We test it out with a three-leg 
ladder with $100$ rungs. The results are shown in Fig.~\ref{threearm}
and, are self explanatory, in the light of the discussions for the 
two-leg case.

\section{Conclusion}

In conclusion, we have shown analytically that an infinite multi-leg 
optical lattice with quasiperiodic modulation in the potential profile 
and in the inter-arm hopping integral can exhibit multiple mobility edges 
if one engineers the phase difference of the modulation 
profiles appropriately by tuning the coherence of the trapping laser beams. 
Exact results corroborated by numerical analysis are discussed specifically 
for a double legged ladder which can be thought to mimic asymmetric double 
well potentials and ultra-cold atoms trapped in higher orbital bands in an 
optical lattice. The results can stimulate experiments to verify the very 
fundamental issues like Anderson localization, metal-insulator transitions 
etc, and may turn out to be useful in designing nano devices.

\section{Acknowledgments}

A.C. thanks a DST-PURSE grant through the University of Kalyani, and 
Atanu Nandy for one of the graphics.

\end{document}